\documentclass[aps, prd, preprint, preprintnumbers, showpacs,
showkeys, amsfonts, nofootinbib, floatfix,11,superscriptaddress]{revtex4-1}
\usepackage{graphics,amssymb,epsfig,float,hyperref}
\usepackage[usenames,dvips]{color}
\usepackage{graphicx}% Include figure files
\usepackage{epsfig}% Include figure files
\usepackage{rotating}% Include figure files
\usepackage{dcolumn}% Align table columns on decimal point
\usepackage{bm}% bold math
\usepackage{amsmath}
\hypersetup{colorlinks,citecolor= nicegreen,linkcolor= nicered}
\definecolor{nicered}{rgb}{0.7,0.1,0.1}
\definecolor{nicegreen}{rgb}{0.1,0.5,0.1}

\def\lsim{\;\raise0.3ex\hbox{$<$\kern-0.75em\raise-1.1ex\hbox{$\sim$}}\;}
\def\gsim{\;\raise0.3ex\hbox{$>$\kern-0.75em\raise-1.1ex\hbox{$\sim$}}\;}
\def\beq{\begin{equation}}   \def\eeq{\end{equation}}
\def\ba{\begin{array}}       \def\ea{\end{array}}
\def\bea{\begin{eqnarray}}   \def\eea{\end{eqnarray}}

\newcommand{\svgg}{\langle \sigma v \rangle_{\gamma \gamma}}

\newcommand{\comment}[1]{}
\newcommand{\newc}{\newcommand}
\newc{\wt}{\widetilde}
\newc{\ra}{\rightarrow}
\newc{\s}{\smallskip}
\newc{\non}{\nonumber}
\begin{document}
\renewcommand{\thefootnote}{\fnsymbol{footnote}}
\vspace{-0.5cm}
\preprint {RECAPP-HRI-2014-002}
%hep-ph/
\vspace{1.2cm}

\begin{titlepage}
\title{Right Sneutrino Dark Matter and a Monochromatic Photon Line}

%\vspace{0.5cm}

\author{Arindam~Chatterjee}
\email{Email: arindam@hri.res.in}
\affiliation{Regional Centre for Accelerator-based Particle Physics, Harish-Chandra Research Institute,  Chhatnag Road, Jhusi, Allahabad 211019, India}
\author{Debottam~Das}
\email{Email: debottam.das@physik.uni-wuerzburg.de}
\affiliation{Institut f\"ur Theoretische Physik und Astrophysik, Universit\"at W\"urzburg,
Am Hubland, 97074 W\"urzburg, Germany}
\author{Biswarup~Mukhopadhyaya}
\email{Email: biswarup@hri.res.in}
\author{Santosh~Kumar~Rai}
\email{Email: skrai@hri.res.in}
\affiliation{Regional Centre for Accelerator-based Particle Physics, Harish-Chandra Research Institute,  Chhatnag Road, Jhusi, Allahabad 211019, India}

%\date{\today}

\vspace{1cm}
\begin{abstract}
The inclusion of right-chiral sneutrino superfields is a rather straightforward
addition to a supersymmetric scenario. A neutral scalar with a substantial
right sneutrino component is often a favoured dark matter candidate in such
cases. In this context, we focus on the tentative signal in the form of a
monochromatic photon, which may arise from dark matter annihilation and
has drawn some attention in recent times. We study the prospect of such a
right sneutrino dark matter candidate in the contexts of both MSSM and NMSSM
extended with right sneutrino superfields, with special reference
to the Fermi-LAT data.
\end{abstract}

%%%%%%%%%%%%%%
\pacs{} 
\keywords{Supersymmetry, Dark Matter, Right sneutrino}
%%%%%%%%%%%%%%
%\vspace*{-0.9cm}
\maketitle
%%%%%%%%%%%%%%
\end{titlepage}

\section{Introduction}
Various observations ranging from galactic rotation curves to the observed 
anisotropy in the cosmic microwave background radiation, have strengthened 
our belief in a substantial cold Dark Matter (DM) component of the universe. It 
is therefore hardly surprising that, side by side with direct searches, all 
indirect evidences of dark matter are also of great interest. 
 
Analyses of the publicly available Fermi-LAT~\cite{Atwood:2009ez} data
found a tentative hint through an observation of a $\gamma$-ray line at $\sim
130$~GeV coming from the vicinity of the galactic center
~\cite{Bringmann:2012vr, Weniger:2012tx, Tempel:2012ey,Su:2012ft}.
In ref. \cite{Weniger:2012tx} it was shown that a possible explanation 
for such a $\gamma$-ray line could be via DM pair annihilations into
two photons, with a DM mass of \mbox{$129.8\pm 2.4 ^{+7}_{-13}$~GeV}
and annihilation cross-section {$\langle \sigma v \rangle_{\gamma\gamma} =
\left(1.27\pm 0.32^{+0.18}_{-0.28}\right) \times 10^{-27} \mbox{cm}^3 
\mbox{s}^{-1}$}  \cite{Tempel:2012ey,Su:2012ft,Hektor:2012kc,Su:2012zg,
Hooper:2012qc, Mirabal:2012za, Hektor:2012jc, Zechlin:2012by}.
Moreover, there is a faint indication ($1.4\sigma$) of 
two lines which can be extracted from the Fermi data, one at $\sim130$~GeV and a
weaker one at $\sim114$~GeV ~\cite{Rajaraman:2012db,Su:2012ft}. Such a pair of
lines can be naturally explained by a DM particle of mass
\mbox{$\sim130$~GeV} annihilating into $\gamma \gamma$ and $\gamma Z$ 
with a relative annihilation cross-section $\langle \sigma v \rangle_{\gamma
Z}/\langle \sigma v
\rangle_{\gamma\gamma}=0.66^{+0.71}_{-0.48}$~\cite{Bringmann:2012ez}.
Though there is no well-accepted astrophysical process that can explain the
$\gamma$-ray line, doubts have been raised \cite{Profumo:2012tr,Boyarsky:2012ca}
to its line feature by making it compatible with a 
diffuse background. 
Preliminary analysis of the Fermi-LAT Collaboration also confirms a line feature
around $\sim 133$ GeV but with a lower statistical significance~\cite{FERMI4}
and concluded that more data would be required to establish the origin of 
such a feature. 

Following reference \cite{Weniger:2012tx}, various models have been
proposed to explain the monochromatic feature of the $\gamma$-ray, see
e.g.~\cite{Cline:2012nw,Lee:2012bq,Dudas:2012pb,Choi:2012ap,
Kyae:2012vi,Acharya:2012dz, Buckley:2012ws,Chu:2012qy,Weiner:2012cb, 
Heo:2012dk, Frandsen:2012db, Park:2012xq,Tulin:2012uq,
Cline:2012bz,Bergstrom:2012bd, Weiner:2012gm, Fan:2012gr,Lee:2012wz,Wang:2012ts,
D'Eramo:2012rr, Bernal:2012cd,Farzan:2012kk,Bai:2012qy,Das:2012ys,
Buchmuller:2012rc,
Cholis:2012fb,Cohen:2012me, Huang:2012yf,Chalons:2012xf,Basak:2013eba,
Tomar:2013zoa,Basak1}. Most models are further constrained from the continuum  
flux of photons arising from annihilations of the DM into $W$ and $Z$ 
bosons and the Standard Model (SM) fermions                                  
\cite{Buckley:2012ws,Buchmuller:2012rc,Cholis:2012fb,Cohen:2012me,
Huang:2012yf}. 
%In order to check whether these constraints are also 
%satisfied, one can define a parameter $R^{\rm{th}}$ \cite{Cohen:2012me} given by
%  
%\begin{equation}
%R^{\rm{th}} \equiv \frac{\langle \sigma v\rangle_{\rm{ann}}}{2\langle \sigma v
%\rangle_{\gamma\gamma}+\langle \sigma v\rangle_{\gamma Z}}, 
%\end{equation}
%where $\langle \sigma v\rangle_{\rm{ann}}$ and $\langle  
%\sigma v\rangle_{\gamma\gamma,\gamma Z}$ are the total annihilation
%cross-section and the annihilation cross-sections into $\gamma\gamma$ 
%and $\gamma Z$ respectively. Assuming the typical total annihilation
%cross-section for the thermal dark matter (\mbox{$\svann \sim 3
%\times 10^{-26}$ $\rm{cm}^3\rm{s}^{-1}$}), the requirements to 
%explain the lines in the Fermi-LAT data ($\svgg \sim 1.2
%\times 10^{-27}$ $\rm{cm}^3\rm{s}^{-1}$ and $\svgZ/\svgg=0.7$)  imply 
%that $R^{\rm{th}} \sim 9$ ~\cite{Cohen:2012me}. A scenario with 
%$R^{\rm{th}} \sim 9$ is found to be compatible with all constraints coming from
%various indirect searches for dark matter. 
Supersymmetric (SUSY) theories with a viable cold DM candidate have been 
well studied in this context. However it is found that within the
minimalistic versions where the lightest neutralino is the DM
candidate, it is very difficult to accommodate the  $\gamma$-ray 
line signal. We can summarize the shortcomings as follows: 
\begin{itemize} 
\item In the minimal version, {\it viz.} the Minimal Supersymmetric Standard
Model (MSSM), it is difficult to obtain the large annihilation cross-section 
of neutralino pairs into photons $\svgg$~\cite{Boudjema:2005hb}, while satisfying 
constraints imposed by the thermal relic density and large
continuum flux~\cite{Acharya:2012dz} data.

\item In the Next-to-Minimal Supersymmetric Standard Model (NMSSM), which
addresses the $\mu$-problem in MSSM, one can accommodate the large annihilation 
cross-section of the neutralino pairs into photons by exploiting the {\it very} 
singlet like CP-odd Higgs boson resonance \cite{ferrer:2006hy, Chalons:2011ia, Das:2012ys}. 
However, the parameter space is tightly constrained by the direct searches for
the DM, most importantly by the data from XENON100 and LUX 
\cite{Aprile:2012nq,LUX13}.  
It is however observed that in a specific region of the parameter space, where
$\mu_{eff}<0$, constraints from direct detection can be relaxed by an order,
in compliance with the bound from XENON100 \cite{Chalons:2012xf}.
\end{itemize}
In addition, a 130 - 135 GeV photon signal can be produced both in the
the MSSM and the NMSSM through internal bremsstrahlung~\cite{Shakya,
Tomar:2013zoa}, although a
a significant boost factor is required \cite{Tomar:2013zoa} in the latter
scenario.
%,Kang:2012bq
%*****************************************************************************

In this work we study the feasibility of a new DM candidate, 
{\it viz.} a right-chiral sneutrino, $\tilde \nu_R$ (with some degree of mixing with a 
left-chiral one), as the candidate for producing the photon line. 
The simplest way to accommodate non-zero
(Dirac) neutrino masses in SUSY models is by introducing a right-handed singlet neutrino. This would 
entail addition of right-chiral neutrino superfields in the 
MSSM\footnote{In the standard seesaw extensions of MSSM,  Majorana mass scale for
the right handed neutrino superfields is very close the gauge coupling
unification scale ($M_G \sim 10^{16}$ GeV) which makes right handed sneutrinos
very massive (close to $M_G$), thus not suitable for electro-weak scale dark
matter candidate.}. In addition we ensure that the 
thermal relic density is in agreement with WMAP data \cite{WMAP} 
and also satisfies constraints from DM-nucleus scattering
\cite{Aprile:2012nq,LUX13}. Thanks to their singlet nature, the right-handed
sneutrinos ($\tilde \nu_R$), acting as cold DM candidates 
\cite{Moroi1,Moroi2,shri}
in the MSSM, can account for all tentative evidences of DM we have so far. 
A sizeable volume of work has also taken place on the LHCs signals 
of (right) sneutrino DM \cite{collider1}, and also on the related scenarios 
carrying implications on different aspects of phenomenology \cite{collider2,DeRomeri:2012qd}. 
However, as we will discuss, a 130-135 GeV $\tilde \nu_R$ DM, 
that can produce $\svgg \sim 1.2 \times 10^{-27}$, falls short in accounting 
for all the continuum constraints.

To get around this difficulty, we consider a similar scenario in the
Next-to-Minimal Supersymmetric Standard Model ($\tilde{\nu}_R$NMSSM) with a
scale invariant superpotential, assuming a $\tilde \nu_R$ type DM. Notably, 
$\tilde \nu_R$ can naturally acquire a Majorana mass term of $\mathcal O(1)$~TeV. In addition 
to contributing neutrino masses and mixings \cite{Kitano-2001,Das_nu},  
$\tilde \nu_R$ DM in the NMSSM may have rich phenomenology as discussed 
in ref: \cite{Cerdeno:2008ep,Cerdeno:2009dv,Cerdeno:2011qv,Huitu:2012rd}. 
As will be discussed in detail, in this case,  one can even evade the continuum 
constraints when considering resonant annihilation mediated via a singlet-like
Higgs boson. Additionally, the constraints  from XENON100 and LUX on the  spin-independent 
direct detection cross--section can also be satisfied. 

This paper is organized as follows: in Section \ref{bwr} we discuss 
resonant annihilation of DM; following which, in Section 
\ref{nurnmssm}, we explore the possibility of explaining the observed 
$\gamma$ signal with $\tilde{\nu}_R$ DM in the MSSM and the NMSSM. 
Finally, we summarize in Section \ref{conclusion}. 
%**************************************************************************
\section{Breit-Wigner Resonance: Effect of threshold}
\label{bwr}
The pair annihilation of $\tilde{\nu}_R$ into two photons can proceed
via a dominantly singlet/doublet like CP-odd (CP-even) Higgs 
$A~ (H)$ in the s-channel.  
%(in case of neutralinos see 
%\cite{ferrer:2006hy, Chalons:2011ia}, see Fig.~\ref{fig:1}). 
\begin{figure}[ht!]
\begin{center}
\includegraphics[scale=0.5,angle=-0]{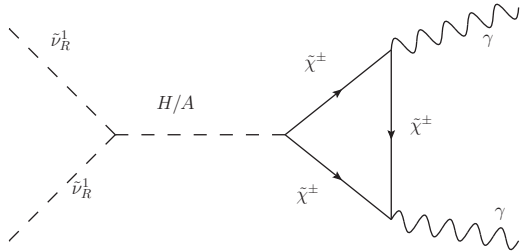}
\end{center}
%\vspace*{-5mm}
\caption{The dominant annihilation diagram for $\tilde{\nu}_R$--like DM
into two photons via a singlet--like CP-even/CP-odd Higgs $H/A$ in the NMSSM. 
Similar diagrams with (s)quarks and (s)leptons running in the loop contribute 
negligibly.}
\label{fig:1}
\end{figure}
Before getting into specific models, we first discuss the
annihilation of a spin-0 dark matter particle ($\tilde{\nu}_R$ in our context)
with mass $m$, mediated via a spin-0 particle of mass $M$ near the resonance.
DM annihilations near resonances and thresholds have been previously
studied \cite{gelmini, murayama, guo}. Our discussion closely follows 
\cite{murayama,guo}. 

%  \frac{s}{M^2} 
The cross--section of an s--channel scattering process, near the resonance,  
is given by, 
\begin{eqnarray}
\label{xsection}
 \sigma  = \frac{32 \pi}{4 E_1 E_2 v \overline {\beta_i}}
\frac{M^2 \Gamma^2}{(s-M^2)^2 + M^2 \Gamma^2} B_i B_f, 
\end{eqnarray}
where $\overline {\beta_i}=\sqrt{1 - \frac{4 m^2}{M^2}}$; $B_i$ and $B_f$ 
are the branching fractions of the intermediate particle into the initial and
final channels respectively and $\Gamma$ is the total decay width of the same;
$E_1$ and $E_2$ are the energies of the two annihilating particles; $s = (p_1
+p_2)^2 $ where $p_1$ and $p_2$ represent the four-momenta of the two
annihilating particles. In the thermal averaging of $\sigma v$, in the context
of DM, the M\o{}ller velocity ($v$) is used \cite{gelmini}. However, 
in the rest frame of one of the annihilating particles, and also in the center
of momentum (CM) frame, the M\o{}ller velocity is reduced to the relative
velocity of these particles. Following Ref. \cite{murayama}, to quantify the
resonance, an auxiliary
parameter $\delta$ is
introduced, such that, 
\begin{eqnarray}
\label{delexp}
M^2 = 4 m^2 (1- \delta) \; . 
\end{eqnarray}
Since, in the present context, the
annihilating particles in resonance are non-relativistic, $|\delta| \ll 1$ is
assumed. Note that for $\delta < 0$, a physical pole $(s = M^2) $ is
encountered when $v \simeq 2 \sqrt{|\delta|}$ (in the CM frame, where
$v$ denotes the magnitude of the relative velocity of the annihilating
particles); while, for $\delta > 0$, a physical pole is never encountered. 
In the former situation $B_i,~ B_f$ are well--defined, consequently 
Eq. (\ref{xsection}) holds good in this region. However, in the latter 
($\delta > 0$), the intermediate particle can no longer decay into two dark
matter particles; consequently $B_i$ and $\overline {\beta_i}$ are unphysical
(imaginary numbers). However, $B_i/\overline{\beta_i}$ remains well-defined. In
this case, Eq. \ref{xsection} can also be expressed as, 
\begin{eqnarray}
\label{belpole}
 \sigma = \frac{2 |C|^2}{4 E_1 E_2 v}
\frac{M \Gamma}{(s-M^2)^2 + M^2 \Gamma^2} B_f \;,
\end{eqnarray}
where, $C$ denotes the coupling between dark matter particles and the mediating
particle which in our context are $\tilde{\nu}_R$ and the CP-even or CP-odd Higgs respectively. 
In the CM frame, with non-relativistic dark matter particles ($v \ll 1$), we have 
$s \equiv 4m^2 (1 + v^2/4) $. Eq. \ref{xsection}, then, reduces to \cite{murayama}, 
\begin{eqnarray}
\label{delgam}
 \sigma v= \frac{32 \pi} {M^2} \frac{\gamma^2 B_f}{(\delta + v^2/4)^2  
+ \gamma^2} \Big(\frac{B_i}{\overline{\beta_i}}\Big),
\end{eqnarray}
where $\gamma = \Gamma/M$.

In the limit of narrow width resonance $\gamma \ll 1$ while for $\text{max}~(|\delta|, 
\gamma) < v \ll 1 $ the denominator of Eq. \ref{delgam} receives dominant 
contribution from $v^2$, and thus $\sigma v$ is enhanced as $v$ decreases. 
This behavior continues until $v \lsim \text{max}~(|\delta|, \gamma)$. Finally,
for 
$v \ll \text{max}~(|\delta|, \gamma)$, $\sigma v$ becomes insensitive to $v$, and is 
then determined only by $\delta$ and $\gamma$. On the contrary, in case of broad
width resonance ($\gamma \gg 1$), with $ \gamma \gg \text{max}~(|\delta|,
v^2/4)$ and $v
\ll 1$), $\gamma^2$ dominates in the denominator, and thus
$\gamma^2/((\delta + v^2/4)^2 + \gamma^2) \simeq 1 $. 

Since DM annihilation into $\gamma \gamma$, in our case, is a
loop--suppressed process, the required $\langle 
\sigma({\tilde{\nu}_R\tilde{\nu}_R}\to \gamma 
\gamma)v\rangle \simeq 10^{-27}\ \text{cm}^3\ \text{s}^{-1}$ apparently leads to
a larger $\langle \sigma_{ann} v\rangle$, where $\sigma_{ann}$ denotes the
total annihilation cross--section of the DM into the SM particles. As all the 
tree-level processes are mediated by the same intermediate state (at resonance), it leads to 
a much lower thermal relic abundance. We therefore now discuss about how we  achieve  
the required cross--sectio $\langle \sigma v \rangle$
for the $\gamma$ signal, as well as the correct relic abundance.

\begin{itemize}
 \item In the context of the $\gamma$ signal, $\langle \sigma v \rangle $ 
annihilation cross--section needs to be evaluated at late times, i.e. 
typically when $v_{rel} \sim 0.001 $. Thus, in Eq. \ref{delgam}, with
$|\delta|\sim \mathcal{O}(10^{-2})$, $v$ can be ignored in the denominator. 
As we shall discuss in the next section, by suitably choosing the coupling $C$
(as in Eq. \ref{belpole}) along with $\delta,\gamma$, it is possible 
to achieve the required cross--section in the $\tilde{\nu}_R$NMSSM. 

 \item 
During freeze-out, away from a pole, the typical velocity of cold DM is about 
$0.3$. Here, by freeze-out, we mean when $(n-n_{eq}) \simeq
n_{eq}$; $n$ and $n_{eq}$ represent the DM density at a given time/temperature 
and
the equilibrium value of the same respectively. Let the corresponding
freeze-out temperature be denoted by $T_f$. \footnote{ We have used 
\texttt{micrOMEGAs} to obtain the freeze-out temperature, and also the relic 
abundance. Note that, \texttt{micrOMAGAs} uses $n(T_f) = 2.5 ~n_{eq}(T_f)$ to 
estimate the freeze-out temperature $T_f$\cite{micrO}.} Since  after freeze-out 
(in the absence of a pole) annihilations do not affect the relic abundance of DM
significantly, the abundance at $T_f$ usually provides a good estimate
of relic abundance. Since DM is non-relativistic at freeze-out, the thermal
abundance at $T_f$ is exponentially suppressed by a factor $x_f =
\dfrac{m}{T_f}$.  
The situation is different for the two regions in the vicinity of the pole, 
namely, $\delta>0$ and $\delta<0$. 

\subitem{$\square$}
When $\delta >0$ (a scenario consistent with narrow width resonance), in the
region $v^2> 4 ~\text{max}~(\delta,~\gamma$), 
the cross--section 
$\sigma$ is dominated by $\gamma/v^2$. Thus, $\langle \sigma v\rangle$, 
at a temperature $T \sim m v_{0}^2 $, $v_{0}^2 \gg 
\text{max}~(\delta,~\gamma)$, 
is determined by $\gamma/v_0^2$. At such large $v_0$, $s > M^2$, and thus
annihilations dominantly occur further away from the pole, and may have 
cross--sections similar to the other annihilation channels (if allowed) not
mediated via the resonance. However, as $T$, and therefore $v_0$, decreases, 
the annihilation channels mediated via the resonance tend to have larger 
$\langle \sigma v\rangle$, and thus, do not decouple. Consequently 
DM can continue to annihilate through these channels until $v_0^2 < 4
~\text{max}~(\delta,~\gamma)$. After that, the corresponding $\sigma$ does not
change any more, and, assuming that these are the only annihilation channels, the 
relic density would be determined by $\delta$ and $\gamma$ only.
\cite{murayama}. 
  
\subitem{$\square$}
For $\delta <0$, the pole is physical. Therefore, unlike the previous 
case, at $s = M^2$ or $v^2 \simeq 4 |\delta|$, $\sigma$ is very large.
At high temperatures, $T \sim m |\delta|$, $\langle \sigma v
\rangle$ is large, and then decreases with $T$. Thus, in this case, 
the annihilation channels mediated via the resonance decouple early. 
Consequently, the relic density of the DM, at late times, remain 
similar to a scenario where DM has a similar annihilation cross-section even 
without the resonance \cite{murayama}. 
For smaller values of $|\delta|$ and $\gamma$ a large boost factor
(defined as $\frac{\langle \sigma v\rangle|_{T \rightarrow 0}}{\langle \sigma
v\rangle|_{T\sim T_f}}$) can be obtained in this case too \cite{guo}. 

\item
The time of freeze-out, while a little away from the pole, also has a moderate 
%, defined by $\Gamma \simeq H$, where $\Gamma$ and $H$ 
%denote the interaction rate with the thermal soup and the Hubble expansion
%rate respectively
dependence on the coupling $C$ (see Eq. \ref{belpole}). When $\delta >0$, for
large $v_0$, such that $v_0^2 \gsim 4~ \text{max}(\delta,~\gamma)$ the
contribution from the resonant 
annihilation can be small compared to $H$, and freeze-out can happen early 
compared to a similar scenario (i.e. with similar value of $C$) with $\delta
<0$. In the former scenario, 
the resonant channel decouples much late (i.e. when $\Gamma \leq H$, where
$\Gamma$ and $H$ denote the interaction rate with the thermal soup and the
Hubble expansion rate respectively) as its contribution
continues to increase until $v_0^2  \lesssim 4 ~\text{max} (\delta,~\gamma)$.
But this late decoupling does not lead to exponential suppression (by $x_D =
\dfrac{m}{T_D}$, where $T_D$ is the decoupling temperature) to the relic
density. On the 
other hand, large annihilation cross--section at high $v_0$ leads to large 
contribution to the thermally averaged annihilation cross-section at early 
times for $\delta <0$. This, in turn, results in late freeze-out and a lower 
relic abundance \cite{murayama}. As we will elaborate in the next section, 
for our benchmark points, we could obtain large relic in the former scenario, 
i.e. with $\delta >0$. On the other hand, since DM annihilations into two
photons happens at late time, even with $\delta >0$ it is possible to obtain 
the required $\svgg$.

\end{itemize}

\section{Photon signal with $\tilde{\nu}_R$ dark matter}
\label{nurnmssm}
In the following, we discuss the different avenues of indirect detections for a
relatively light $\tilde{\nu}_R$ dark matter focusing on the $\gamma$ ray line
observed
at $E_{\gamma}\sim 130-135$ GeV. As mentioned earlier, we extend the MSSM and the 
NMSSM  with three generations of right handed neutrino superfields ($\hat{\nu}_R^c$).
Assuming that the $\tilde{\nu}_R^c$ is the lightest SUSY particle (LSP) in the
supersymmetric particle spectrum, we enforce that the following
phenomenological constraints always hold.
\begin{itemize}
\item A relic density complying with the WMAP bound $\Omega h^2 =
0.1120\pm 0.011$ \cite{WMAP} (with $2 \sigma$ error bars).
\item A SM-like Higgs boson with $M_{H_{SM}} = 124 - 127$~GeV.
\item Constraints from B-physics which has little impact for the 
$\tan\beta$ considered here.
\item Upper bounds on annihilation cross sections into $W^+ W^-$, $Z Z$,
$b\bar{b}$ and $\tau \bar{\tau}$ channels from the Fermi~LAT
collaboration \cite{Ackermann:2012qk,Ackermann:2011wa}, as well as bounds 
from PAMELA on the anti-proton flux \cite{Adriani:2010rc}. 
\end{itemize}

For exact calculation of the sneutrino mass and mixing matrices as well as 
two-loop renormalization group equations (RGEs) for all SUSY parameters,  
we have used the publicly available code called \texttt{SARAH} \cite{SARAH}. These
RGEs are then implemented in the software package \texttt{SPheno} \cite{SPheno} 
for numerical evaluation of all physical parameters and phenomenological constraints.   
For computation of relic density, all indirect detection cross--sections 
and fluxes, we implement \texttt{SARAH} generated \texttt{CalcHEP} \cite{chep} 
model files into \texttt{micrOMEGAs} \cite{micrO}. We however calculate the cross-section
for the photon line $\svgg$ signal with our own mathematica code based on 
Ref. \cite{murayama,guo}. We ignore the contribution of $\svgg$ in the 
relic density computation.

In both the models that we have considered, 
we scan the parameter space while keeping the soft 
SUSY breaking terms in the following preferred ranges.

\begin{itemize}
\item Squark masses of 2-3~TeV are assumed
to alleviate LHC constraints from direct SUSY
searches. The latter choice also helps to enhance the lightest Higgs boson mass
irrespective of the choice of $\tan\beta$. Similarly, gluino masses ($M_G$) is
fixed around $\sim$ 2 TeV. The slepton masses are assumed to be around
~300 GeV to have consistent spectra with muon anomalous magnetic moment.
\item  Trilinear soft susy breaking terms $T_t = -3$~TeV and $T_b = -1.0$~TeV
(scaled with the Yukawa couplings). 
\item  We use $m^2_{\tilde{\nu}_R}$ as free parameter. Similarly,
the couplings $y_\nu$ and $T_\nu$ for $\nu_l-\nu_R$ and
$\tilde\nu_l-\tilde\nu_R$ are assumed flavor diagonal. In the present context,
we refrain from exact calculation of neutrino masses and mixing angles.  
\item $\mu \sim \pm (200-300)$ GeV is assumed. 
%Note that, negative $\mu$ is
%somewhat disfavored concerning the discrepancy in the $\Delta a_{\mu}$. 
\item We have used the top quark pole mass $m_\text{top}$ = 173.1~GeV.
\end{itemize}

\subsection{$\tilde{\nu}_R$ and the MSSM}  
In this section, we discuss the status of $\tilde{\nu}_R^c$ dark matter in the
(R-parity
conserving) MSSM with three generations of right-handed (s)neutrinos. 
The neutrino masses arise from the Yukawa interaction only ({\it purely Dirac-type}) and 
can be obtained from the following superpotential:
\begin{eqnarray}
\label{eq:SP}
  W  = W_{MSSM} + y_\nu \, \hat H_u \cdot \hat L \, \hat \nu_R^c 
\, ,
\end{eqnarray}
where $W_{MSSM}$ denotes the MSSM superpotential; $\hat H_u$, $\hat L$ 
and $\hat \nu_R^c$ represents the up-type Higgs, lepton doublet and
right-handed 
neutrino superfields respectively. For simplicity, we consider all mass and 
coupling parameters to be real and suppress flavor indices for neutrino
families.
Assuming soft-supersymmetry breaking, as in the MSSM, the soft-breaking scalar 
potential becomes,   
\begin{eqnarray}
\label{eq:Soft}
  V_{soft} 
  = V_{MSSM} + (T_\nu \,H_u \cdot \tilde L \,\tilde \nu_R^c + \text{h.c.})+ 
  m_{\tilde \nu_R}^{2}|
\tilde \nu_R^c|^2\, ,
\end{eqnarray} 
%add generation index i in Yukawas and soft terms
where $V_{MSSM}$ denotes the soft-supersymmetry breaking terms in the MSSM and
soft trilinear coupling is given by $T_\nu \equiv 
T_{\nu_\alpha}y_{\nu_\alpha}$, where $\alpha$ denotes the generation index.
Neutrino masses can be expressed as,  
\begin{eqnarray}
  m_\nu = y_\nu \, \langle H_u^0 \rangle
  = y_\nu \, \frac{v}{\sqrt{2}} \, \sin \beta \, 
\end{eqnarray}
where $v \simeq 246$ GeV is the vacuum expectation value (VEV) of the
standard-model
-like Higgs boson, and $\tan \beta = \langle H_u^0 \rangle / \langle 
H_d^0 \rangle$.  Clearly, neutrino masses ($m_\nu \sim 0.1$ eV) put constraints
on the size of the neutrino Yukawa couplings $y_\nu \sim \mathcal{O}(10^{-12})$. 
Ignoring flavor mixing in the $\tilde{\nu}$ sector \footnote{Although flavor 
violation cannot be ignored in the $\nu$ sector, in $\tilde{\nu}$ sector, it is
only induced through terms proportional to $y_{\nu}$, assuming the soft-breaking
terms to be flavor conserving. Since $y_{\nu}$ is $\mathcal{O}(10^{-12})$, such
effects are very small and do not affect our results.} the
(2$\times$2) mass matrix for $\tilde{\nu}_{\alpha}$ for any flavor $\alpha$, 
can be written as,  
% mention basis
\begin{equation} 
m^2_{\tilde{\nu}_{\alpha}} = \left( 
\begin{array}{cc}
m^{\alpha 2} &\frac{1}{\sqrt{2}} \Big(- v_d \, \mu \, y^{*}_{\nu_\alpha}  +
v_u T^*_{\nu_\alpha } 
\Big)\\ \frac{1}{\sqrt{2}} \Big(- v_d \, \mu^* \, y_{\nu_\alpha}  + v_u
T_{\nu_\alpha} 
\Big) &\frac{1}{2} v_{u}^{2} |y_{\nu_\alpha}|^2  +
{m^{2}_{\tilde \nu_{R\alpha}}}\end{array} 
\right) 
 \end{equation} 
with
\begin{align} 
m^{\alpha 2} &= \frac{1}{8} \Big(4 v_{u}^{2} |y_{\nu_\alpha}|^2  + 
8 m_{\tilde {l}_\alpha}^{2}  + 
\Big(g_{1}^{2} + g_{2}^{2}\Big)\Big(- v_{u}^{2}  + v_{d}^{2}\Big)\Big),
\end{align} 
where $m_{\tilde
{l}_\alpha}$ is the
soft breaking term for $\tilde {l}_\alpha$. 
The lightest sneutrino mass eigenstates $\tilde{\nu}_R^{1}$, as obtained after 
diagonalization of $\tilde \nu_{\alpha}$, can be a valid candidate for DM. 
Moreover, such a $\tilde{\nu}_R^{1}$ can also produce viable photon signal 
$\langle \sigma v\rangle_{\gamma\gamma,\gamma Z} \sim 10^{-27}
~\text{cm}^3\text{s}^{-1}$, through the resonant annihilations via 
heavier CP-even ($H_2$) or CP-odd Higgs boson ($A$). Note that the resonant
annihilation 
through CP-odd Higgs boson $A$ can only take place if $T_{\nu_{\alpha}}$ is complex, 
as otherwise the coupling among $A \, \tilde{\nu}_R^{1} \tilde{\nu}_R^{1*}$ 
vanishes. The CP-odd Higgs resonance avoids stringent constraints on 
$\langle \sigma v\rangle_{W^{+} W^{-},Z Z}$ coming from continuum fluxes of
gamma rays
\footnote{Note that a complex value for $T_\nu$ leads to a mixing among 
the CP-even and CP-odd Higgses in the mass eigen-basis.}. 
We found that it requires large values of ${\mathcal Im} (T_{\nu_{\alpha}})$ to
obtain 
the  desired cross-section for the di-photon final state. However, a dominantly 
CP-odd Higgs resonance falls short in accounting for the desired thermal relic
abundance; and, more importantly, the viable parameter space has been ruled out
by constraints from the  continuum spectrum of $\gamma$ rays. In particular, 
we found that annihilation cross-sections to the following final states are somewhat
above the upper limits set by the Fermi-LAT for a NFW halo profile 
\cite{Bernal:2012cd,Huang:2012yf}, {\it viz.} (i) $\langle \sigma v\rangle  
(\tilde{\nu}_R^{1} \tilde{\nu}_R^{1} \to b \bar{b}) \sim 10^{-23}\text{cm}^3\ \text{s}^{-1}$ 
and (ii) $\langle \sigma v \rangle (\tilde{\nu}_R^{1} 
\tilde{\nu}_R^{1} \to Z H )\sim 10^{-23}\text{cm}^3\ \text{s}^{-1}$.  

Also, a significant left-right mixing in the lightest sneutrino states,  which enhance the photon signal, 
is tightly constrained from the direct DM searches. Thus, in this extension of the MSSM, although 
$\tilde{\nu}^1_R$ is a viable DM candidate, 
achieving large $\svgg$ seems to be incompatible with other constraints
on DM.  

\subsection{$\tilde{\nu}_R$ and the NMSSM}  
Next we consider a similar extension of the scale invariant 
NMSSM, with three generations of right-handed neutrino superfields.
The new superpotential becomes
\begin{eqnarray}
\label{wnmssm}
  W 
  = W_{MSSM} + \lambda \widehat S \widehat H_u\cdot \widehat H_d +
\frac{\kappa}{3}  \widehat S^3 + y_\nu \, \hat H_u \cdot \hat L \, \hat
\nu_R^c+ 
   \frac{ y_r}{2} \, \hat S \, \hat \nu_R^c \, \hat \nu_R^c
\, ,
\end{eqnarray}
where $W_{MSSM}$ denotes the MSSM superpotential without the $\mu$ term. The $\hat S$ denotes 
the singlet superfield that already appears in the NMSSM. When the scalar component of 
${ \hat S}$ gets a VEV of the order of electro-weak scale, $\mu$ term of correct size would be generated. Similarly, the right handed 
neutrinos also acquire an effective Majorana mass around the electro-weak 
values as long as the dimensionless coupling $y_r$ is order one
\cite{Kitano-2001}. 
At the tree level the $(3 \times 3)$ light neutrino mass matrix, that  arises 
via the seesaw mechanism, has a very well-known structure given by, 
\begin{equation}
\label{seesaw-formula}
m_{\nu}^{tree} = -m_DM^{-1}_Rm^T_D,
\end{equation}
Unlike MSSM, here $y_{\nu} \sim \mathcal{O}(10^{-6})$ can reproduce the neutrino
mass and mixing data \cite{Das_nu}. As before, flavor mixings in the slepton sector can 
be induced radiatively by the off-diagonal entries in the neutrino Yukawa coupling,
which are suppressed due to the smallness of Yukawa couplings. However,
for simplicity we assume
neutrino Yukawa couplings to be diagonal which alleviates the slepton flavor
mixings completely.  
The soft-supersymmetry-breaking scalar potential is given by, 
\begin{eqnarray} 
%\label{eq:SoftN}
\label{vnmssm}
  V_{soft} 
  &=& V_{MSSM} +(
T_\lambda\, H_u \cdot H_d \,S +  \frac{1}{3} T_\kappa\,  S^3
\ \nonumber \\ \nonumber &+& T_\nu \,  H_u \cdot  \tilde L \,  \tilde \nu_R^c + 
    T_{r} \,  S \, \tilde \nu_R^c \, \tilde \nu_R^c + \text{h.c}) 
+ m^{2}_{\tilde \nu_R}|\tilde \nu_R^c|^2 + m^2_{S} |S|^2 \, \, ,
%\label{vnmssm}
\end{eqnarray}
where $V_{MSSM}$ denotes the soft-supersymmetry breaking terms in the MSSM
and  $T_\nu \equiv  T_{\nu_\alpha}y_{\nu_\alpha}$ where $\alpha$ denotes the
generation indices. We do not assign VEV to $\tilde{\nu}^c_R$, thus R-parity is
unbroken at the minimum of the scalar potential. In particular, the neutral
scalar fields can develop, in general, the following vacuum expectation values
at the minimum of the scalar potential.
\begin{eqnarray}
\langle H^0_d\rangle &=& v_d /\sqrt 2 ;\hskip 0.2 cm  \langle
H^0_u\rangle =  v_u/\sqrt 2 ; 
\hskip 0.2 cm  \langle S \rangle = v_s/\sqrt 2 \hskip 0.2 cm  \nonumber .
%v_{\tilde \nu_i}&=&\langle \tilde \nu_i \rangle; \hskip 0.2 cm
%v_{\tilde \nu^c_R}= \langle \tilde \nu^c_R \rangle.
\end{eqnarray}
\noindent
We further assume that $y_r$ and $T_r$ are flavor-diagonal for simplicity and consider the lightest 
$\tilde{\nu}^c_{R}$ as the lightest R-parity odd particle; and forefore 
the DM candidate.  

We begin by decomposing the sneutrino fields in terms of real and imaginary 
components. 
\bea
\label{eq:nulnur}
 \tilde{\nu}_L &=&  \frac{1}{\sqrt{2}}(\phi_L + i \sigma_L), \nonumber \\
 \tilde{\nu}^c_R &=&  \frac{1}{\sqrt{2}}(\phi_R + i \sigma_R). 
\eea
where, $\phi_L, \phi_R $ are the $CP$-even and $\sigma_L, \sigma_R$ are the 
$CP$-odd scalar fields.
The presence of $\Delta L=2$ terms (for {\it e.g.} $y_r \, \hat S \, \hat \nu_R^c 
\, \hat \nu_R^c$) in the 
superpotential, and the corresponding soft-breaking term 
($T_{r} \,  \tilde S \, \tilde \nu_R^c \, \tilde \nu_R^c + \text{h.c.}$)  
can induce splitting between different CP-eigenstates of $\tilde{\nu}^c_{R}$. 
 
The mass matrix for CP-even eigenstates  
$\left(\phi_L, \phi_R \right)$ or for CP-odd eigenstates  
$\left(\sigma_L, \sigma_R \right)$ is given by, 

\begin{equation} 
\label{eq:nuI}
m^2_{\nu^{R,I}} = \left( 
\begin{array}{cc}
m^{R,I}_{11} &m^{R,I}_{12}\\ 
m^{R,I}_{21} &m^{R,I}_{22} \end{array} 
\right) ,
 \end{equation} 
where,
%\bea
\begin{align}
\label{eq:nuImat}
&m^{R,I}_{11} &=& \frac{1}{8} (4 v_{u}^{2} y^{2}_{\nu}  
+ 8 m_{\tilde l}^2  + (g_{1}^{2} + g_{2}^{2})(- v_{u}^{2}  + v_{d}^{2})),\nonumber \\ 
&m^{R,I}_{12} &= m^{R,I}_{21} =& \frac{1}{4} (2 \sqrt{2} v_u T_{\nu} 
\pm  v_s (\mp 2 v_d \lambda y_{\nu}  + 4 v_u y_{r}  y_{\nu})), \\ 
&m^{R,I}_{22} &=& \frac{1}{4} (4   m_R^2+ 8   y_r^2v_{s}^{2} +
2 y_{\nu}^{2} v^2_u  \pm
2 \Big(( 2 \kappa v_{s}^{2}-\lambda v_d v_u  
) y_{r} +4 \sqrt{2} T_r v_s  - 2 \lambda y_{r}v_u v_d) \Big). \nonumber 
%\eea
\end{align}
 
The mass matrices can be diagonalized by unitary matrices \(Z^I\) and 
\(Z^R\),
\begin{align} 
\label{eq:nudiag}
& Z^I m^2_{\nu^I} Z^{I,\dagger} = m^{D ~2}_{\nu^I}, \nonumber \\ 
& Z^R m^2_{\nu^R} Z^{R,\dagger} = m^{D ~2}_{\nu^R}, 
\end{align} 
where  $m^{D}_{\nu^I}$ and $ m^{D}_{\nu^R}$ denote the diagonalized 
mass matrices respectively, and the corresponding mass eigenstates are 
denoted by $\sigma_i$ and $ \phi_i$ ($i \in \{1,2\}$); $g_1$ and $g_2$ are the
$SU(2)_L$ gauge couplings; $m_{\tilde l}^2$ is the soft-supersymmetry breaking mass 
term for slepton doublet. The other parameters are described in Eqs.
(\ref{wnmssm}) and (\ref{vnmssm}). Generically, 
$\phi_i$ and $ \sigma_i$ are non-degenerate, thanks to the $\Delta L=2$ 
term present in the superpotential.  

In general, lightest of the states $\sigma_i$ or $\phi_i$ could be the 
lightest SUSY particle in different 
regions of the parameter space. Depending on the choice of the parameters,
$\sigma_i$ or $\phi_i$ can have dominant gauge and/or Yukawa interactions.
Their mass difference, defined by $\Delta m=|m_{\phi_i}-m_{\sigma_i}|$ cannot be
arbitrary, especially when $\sigma_i$ and $ \phi_i$ have dominant left-handed 
component, i.e., $\sigma_i$ and $\phi_i$ are of $\tilde{\nu}_L$-type. In this
case, the one-loop contributions to the neutrino mass matrix can be quite large which
essentially limits $\Delta m \simeq 100$ keV \cite{Haber_nu,Das_nu}. Moreover, 
due to its doublet nature under $SU(2)_L$, stringent
constraints would also appear from the sneutrino-nucleus scattering (via
$t$-channel $Z$ boson exchange processes)  
%if the
%lightest CP eigenstate $\sigma_1$ and/or $ \phi_1$ is the desired candidate 
%for DM 
\cite{Falk:1994es}.

On the other hand, aforementioned constraints can be evaded naturally, if 
we assume that $\phi_1$ and $\sigma_1$ are dominantly right-handed. 
In fact these states are completely unconstrained, and their splitting can be 
traced back to $\Delta L=2$ terms present in the mass matrix. However, 
nearly degenerate or degenerate ${\sigma_1}$ and $ {\phi_1}$ may be achieved, 
provided all $\Delta L=2$ in combination vanishes. 
The condition for degeneracy, thus, can be expressed as, 
\begin{equation}
\label{nudeg}
\Big( \Big(- v_d v_u \lambda  +  v_{s}^{2} \kappa \Big) y_{r}
+\sqrt{2} v_s T_r\Big) = 0. 
\end{equation}
\begin{table}[t!]
\begin{center}
\begin{tabular}{|c|ccc|c|ccc|} \hline
parameter       & A     &  B    & C      & parameter     & A   &  B      &  C 
\\ \hline
$\tan\beta$    & 2.9      &9.0      &  10.0  & $\mu_{eff}$       & 322    &
-273.2  & -269.3 \\
$\kappa$       & 0.712    & 0.713   & 0.713  & $T_{\kappa}$      & -45.14 &
277.9   & 273.92 \\     
$y^{33}_r$     & 0.18     & 0.506   & 0.50   & $T^{33}_r$     & -36.10   &-79.98
  & -80.11  \\
$\lambda$      & 0.719    & 0.73    & 0.719   & $T_{\lambda}$  & 418     
&-1077   & -1077 \\
$m_{\tilde{b_{1,2}}}$     &3200     &3200    &  3200 & $m_{\tilde{Q}}$ &~2000 
&~2000    & ~2000 \\
$m_{\tilde{l}}$    &~300  &~300     &~300    & $m_{H_{1}}$     & 126    &
125.8   & 127  \\
$m_{H_{2}}$    & 332      & 260.14  & 259.92 & $m_{H_{3}}$     & 1893   &1810 
   & 1903\\
$m_{A_{s}}$    & 269.7    & 456.7   &  452   & $m_{A_{3}}$     & 1891   & 1807
   & 1903 \\
$m_{H^{\pm}}$  & 1887     & 1803    &  1897  & $m_{\tilde{\nu^{R}_{1}}}$ & 135 
&304 &  305 \\
$m_{{\wt\chi_1}^{\pm}}$   & 197     & 201.6  &  196 & $m_{\tilde{\nu^{I}_{1}}}$ 
& 135 & 130.6 &  130 \\
$m_{{\wt\chi_1}^{0}}$     & 177     & 179.6  &  179  & $m_{\tilde{t_{1,2}}}$
&3200&~3200    & ~3200 \\ 
\hline
$\sigma$({\small{DM~DM}} $\to \gamma \gamma)$ & 2.0 & 2.0 & 1.3 &
$\Omega_{\phi_1,\sigma_1}h^2$ & 0.11  & 0.10 & 0.09 \\
$(10^{-27}\ \text{cm}^3\ \text{s}^{-1})$ &  &  &  &&&& \\ 
\hline
\end{tabular}
\end{center}\caption{Benchmark points for $\sigma_1$ and/or $\phi_1$ DM
(co-) annihilation via (A) CP-odd Higgs pole; (B) CP-even Higgs pole with 
single component DM; and (C) CP-even Higgs pole with degenerate 
$\sigma_1$, $\sigma_2$ DM. All masses are shown in GeV.}
\label{tabnusm}
\end{table}
Though, it seems a bit fine-tuned, we find that, potentially testable 
photon signals can be achieved. Based on the above facts, we consider 
the following possibilities for a 130-135 GeV $\tilde{\nu}_R$ type 
dark matter, as presented in Table \ref{tabnusm}. 
%%%

%%%%
%\begin{table}[t!]
%\begin{center}
%\begin{tabular}{lrccc} \hline
%parameter      && A &  B  & C  \\ \hline
%$\tan\beta$     && 2.9   &10.0   &  10.0  \\
%$\mu_{eff}$    && 322   & -259.0 & -269.3 \\
%$\kappa$         && 0.712 & 0.702 & 0.713  \\
%$y^{33}_r$       && 0.18  & 0.5   & 0.50   \\     
%$T^{33}_r$       && -36.10 &-78.0& -80.11 \\
%$T_{\kappa}$      && -45.14 & 276  & 273.92 \\
%$T_{\lambda}$      && 418    &-1077   & -1077.0\\
%$m_{\tilde{t_{1,2}}}$  &&~3200      &~3200 & ~3200 \\
%$m_{\tilde{b_{1,2}}}$  &&~3200      &~3200 & ~3200 \\
%$m_{\tilde{Q}}$ &&~2000   &~2000 &  ~2000 \\
%$m_{\tilde{l}}$   &&~300    &~300  &  ~300 \\
%$m_{H_{1}^0}$      && 126  & 126  &  127  \\
%$m_{H_{2}^0}$      && 332  & 259  &  259.92\\
%$m_{H_{3}^0}$      && 1893 & 500  &  1903\\
%$m_{A_{2}^0}$       && 269.7& 367  &  452 \\
%$m_{A_{3}^0}$       && 1891 & 486  &  1903 \\
%$m_{H^{\pm}}$      && 1887 & 480  &  1897\\
%$m_{\tilde{\nu^{R}_{1}}}$  && 135  & 202&  305 \\
%$m_{\tilde{\nu^{I}_{1}}}$  && 135  & 130&  130 \\
%$m_{{\wt\chi_1}^{\pm}}$    && 197  & 169&  196  \\
%$m_{{\wt\chi_1}^{0}}$     && 177  & 174&  179 \\
%$\Omega_{\phi_1,\sigma_1}h^2$ && 0.105 & 0.05& 0.09 \\
%$\sigma$({\small{DM~DM}} $\to \gamma \gamma)(10^{-27}\ \text{cm}^3\ \text{s}^{-1})$
%&& 20.0 & 3.0 & 1.3 \\ 
%\hline
%\end{tabular}
%\caption{Benchmark points for $\sigma_1$ and/or $\phi_1$ DM
%(co-) annihilation via (A) CP-odd Higgs pole; (B) CP-even Higgs pole with 
%single component DM; and (C) CP-even Higgs pole with degenerate 
%$\sigma_1$, $\sigma_2$ DM. All masses are shown in GeV.}
%\label{tabnusm}
%\end{center}
%\end{table}

\subsubsection{{\large{\it $\phi_1$/$\sigma_1$ Dark Matter}}}
In the first scenario, we illustrate how CP eigenstates $\phi_1$ or $\sigma_1$ can  
annihilate through singlet-like CP-even Higgs ($H_2$) resonance to 
$\gamma \gamma$ (see Fig. \ref{fig:1}). The singlet nature of $H_2$ 
helps to accommodate constraints from the continuum $\gamma$. In addition, 
the narrow width of a singlet--like $H_2$ also reduces the contribution 
of the resonance mediated channels to the relic density. 

The couplings $C_{r,o}$ in Eq. \ref{belpole}, between the singlet-like 
CP-even Higgs and $\phi_1$/$\sigma_1$ takes the following form, 
%\begin{equation}
\begin{align}
C_r &\simeq   \sqrt{2} T_r Z^H_{23} Z^{R~2}_{12}  + v_s 
\left(2 \kappa y_r+ y_r^2 \right) Z^H_{23} Z^{R~2}_{12}+\ldots  \nonumber \\ 
C_o &\simeq   -\sqrt{2} T_r~ Z^H_{23}~ Z^{I~2}_{12}  - v_s 
\left(2 \kappa y_r - y_r^2 \right) Z^H_{23}Z^{I~2}_{12} + \ldots 
\label{eq:Cro}
\end{align}
%\end{equation}
The term proportional to $T_r$ comes from the soft-supersymmetry breaking 
sector, while the other terms come from the F-term contributions to the 
scalar potential. $Z^H$ represents the mixing matrix of 
the CP-even Higgs bosons. Assuming that the second lightest CP-even Higgs (singlet-like) 
boson contributes dominantly,  $Z^H_{23}$ is the relevant entry in Eq. \ref{eq:Cro} while the 
ellipsis indicate the sub-leading contributions originating from small doublet components.
These couplings play crucial roles in determining both $\svgg$ and the thermal relic abundance.

\begin{figure}[h!]
\begin{center}
\includegraphics[height=2.5in]{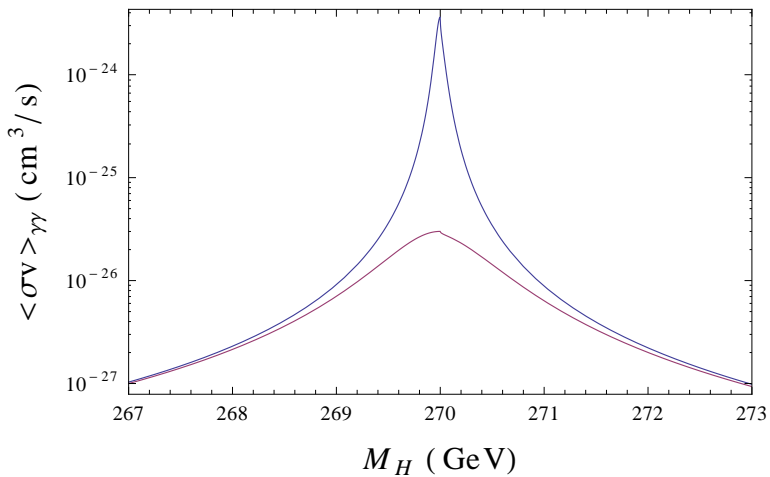}
\end{center}
\caption{The enhancement of the $\langle \sigma v
\rangle_{\gamma \gamma}$ in the vicinity of the resonance, above and below 
the pole (270 GeV) is shown here. The blue curve represents a total
width of 0.1 GeV below the threshold while the red curve represents a width of 1.1 GeV in the
same region. }
\label{fig:pole}
\end{figure}
\begin{figure}[t!]
\begin{center}
\includegraphics[height=2.45in]{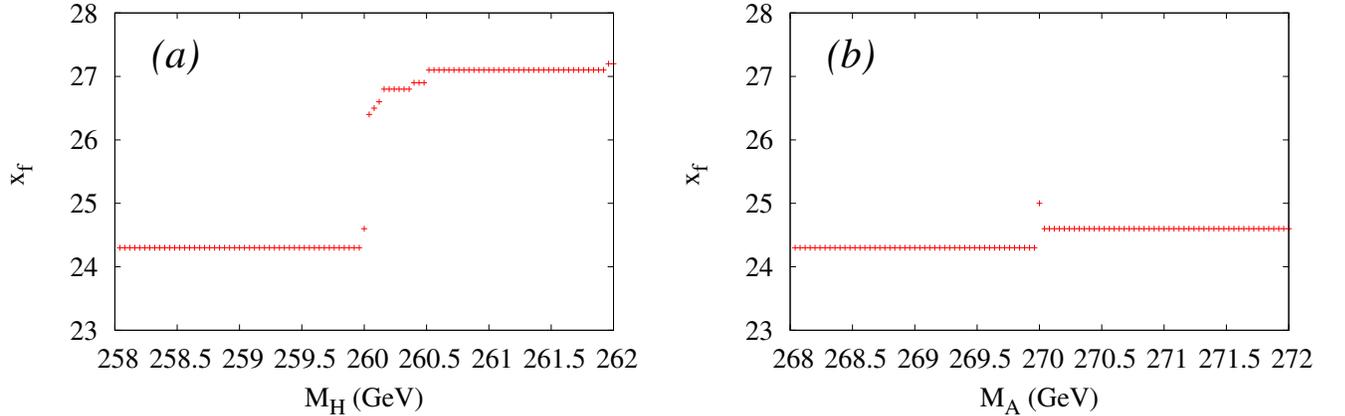}
\end{center}

\caption{$x_f = \dfrac{m_{\sigma_1}}{T_f}$, where $\sigma_1$ is 
dark matter, around the singlet--like ($a$) CP-even Higgs pole and ($b$) CP-odd
Higgs pole have been plotted. Note that, for 
our benchmark points, $m_{\phi_1/\sigma_1}$ is 130 GeV (135 GeV) for
the CP-even (CP-odd) Higgs resonance respectively. This figure depicts
the late freeze-out of the DM when the mass of the DM is less than half of 
the CP-even/CP-odd Higgs mass, i.e. when the physical pole is encountered.}
\label{fig:sv_xf}
\end{figure}

%%%
The desired signal $\svgg$ can be easily enhanced with larger $C_{o/r}$ and/or $\lambda$. 
Larger values of $\lambda$   
can enhance the effective coupling of $H_2-\gamma -\gamma$ via  
$\lambda H_2 (S)\tilde{H_u}^+ \tilde{H_d}^-$ loops in 
Eq. \ref{wnmssm}. Since the contribution from the charginos ($\chi^\pm$) 
running in the loop dominates, light higgsino--like $\chi^\pm$ are desired to 
enhance the signal. In Fig. \ref{fig:pole}, we present $\svgg$ with  
representative values of the input parameters 
around the Higgs threshold. Interestingly, we can easily obtain the $\svgg$
with the pole mass below or above the threshold, i.e. ($2 m_{\sigma_1/
\phi_1}$). 
%%%%
\begin{figure}[h!]
\begin{center}
\includegraphics[height=2.45in]{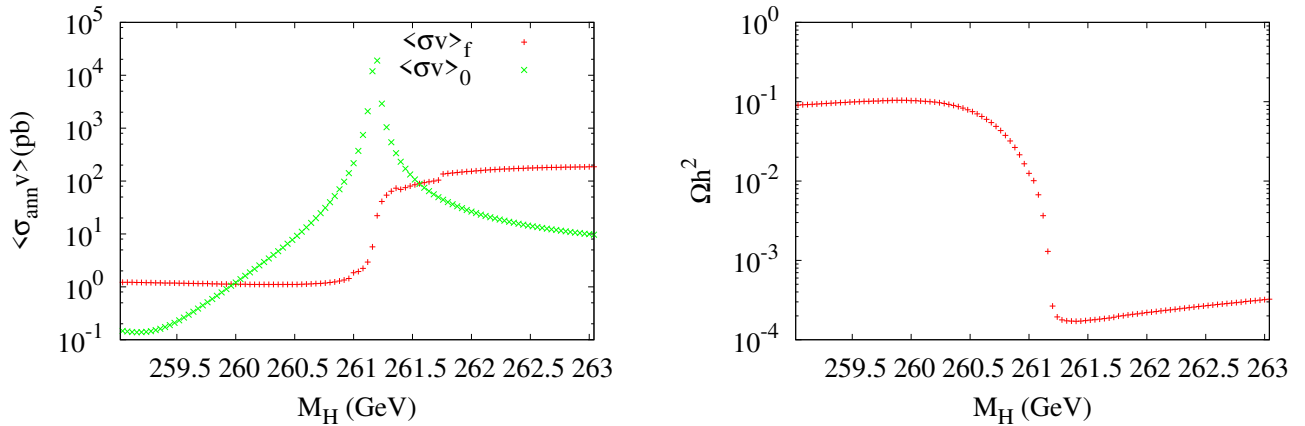}
\end{center}
%~~(a)~~~~~\hspace{2cm}~~~~~~~~~~~~~~~~~~~(b)
\caption{Relic density of $\sigma_1$ DM, with a mass of 130.6 GeV, has been
plotted against the singlet--like CP-even Higgs pole mass. In the left panel
the thermally avaraged cross-sections at freeze-out temperature ($T_f$) and at 
late time (the speed of DM $v \sim 0.001$ or $T=T_0 \sim 10^{-4}$ GeV) have
been shown in red and green respectively. In the right panel 
the CP-even Higgs mass is varied to demonstrate the relic abundance below 
and above the threshold (261.2 GeV). All other relevant parameters have been
mentioned in column (B) of Table \ref{tabnusm}.}
\label{fig:even}
\end{figure}
%%%
However, the correct thermal relic density can be obtained only when the pole 
is below the threshold, as shown in the left panel of Fig. \ref{fig:even}.
To understand it better, we also depict the parameter $x_f$ (i.e. $m/T_f$, as 
already discussed in Sec. \ref{bwr}), which characterises the freeze-out,
against $M_H$ in the
Fig. \ref{fig:sv_xf}. As can be seen from Fig.
\ref{fig:sv_xf}, $x_f$ is larger for $M_H$ above the threshold (260 GeV),
implying late freeze--out compared to the scenario when $M_H$ is below the
threshold. Therefore, in this region 
a lower relic density is obtained. 

\begin{figure}[h!]
\begin{center}
\includegraphics[height=2.45in]{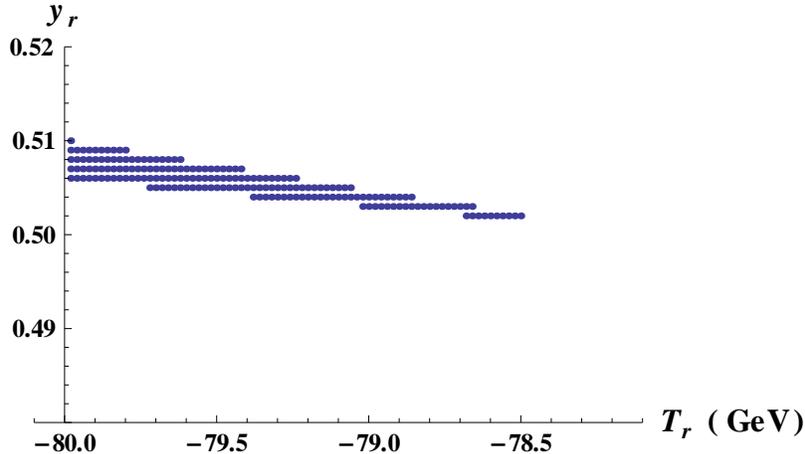}
\end{center}

\caption{WMAP allowed parameter space for $\sigma_1$ DM, with a mass of 130 GeV,
has been shown in the $T_r,y_r$ plane. All these points satisfy direct
detection limits from LUX while having adequate cross-section to account for the
photon signal.}
\label{fig:trfig}
\end{figure}

As noted in Eq. \ref{eq:Cro} and Eq. \ref{eq:nuImat}, both the couplings
$C_{r,o}$, as well as the mass of $\sigma_1$, depend on $T_r$ and $y_r$, 
while the other free parameters involved in these equations also affect, 
among others, the Higgs sector. Therefore, in Fig. \ref{fig:trfig}, we present 
the allowed parameter space in the $(T_r,y_r)$ plane; assuming the following 
input parameters (at the SUSY scale): 
\begin{equation}
\lambda~=~0.73,~\kappa~=~0.713,~T_\lambda~=~-1077,~T_\kappa~=~277.92,
~\tan\beta~=~10. \nonumber
\end{equation}
Our choice for these parameters are consistent with 
the LHC data on the SM-like CP-even Higgs  while providing us with 
another singlet-like CP-even Higgs with a mass of ~260 GeV. 
All other parameters are set to alleviate LHC constraints 
as mentioned in Sec. \ref{nurnmssm}. 
We also set soft breaking parameter of the right handed sneutrino so to have
$\sigma_1$ as LSP with mass $130.5$ GeV. With the given parameters, we obtain
$m_{H_{SM}}=125.7$ GeV while for the singlet like Higgs we get $m_{H_2}=260.1$
GeV. Assuming, for simplicity, that the third generation right handed sneutrino
as the LSP, we only varied $T^{33}_r$ and $y^{33}_r$ to deliniate the
WMAP satisfied region in the vicinity of $H_2$ resonance. In this
region, due to the singlet nature of $H_2$ and the smallness of the
coupling among $\sigma_1-H_2-\sigma_1$, the annihilation channels
mediated by the lightest CP-even Higgs boson contributes singnificantly 
in satisfying the relic density. These points are also allowed by the recent
bounds from LUX~\cite{LUX13}. Interestingly, the whole parameter space can
provide with adequate cross-section for $\svgg$. \\

For the benchmark point presented in column (B) of Table \ref{tabnusm}, we 
assume $\sigma_1$ to be the DM.\footnote{Note that 
by simply reversing the signs of $y_r$ and $T_r$, one can have CP-even
$\phi_1$ as the LSP while the mass spectra remains unchanged. Consequently,
one can explain the observed $\gamma$ signal with CP-even $\phi_1$ DM.} 
The representative values, as shown in column (B) of Table \ref{tabnusm} 
satisfy the desired $\svgg$ and other mentioned constraints including WMAP. 
In the calculation of $\svgg$, we have $\rm Br(H_2 \rightarrow \gamma \gamma)
=1.5\times 10^{-4}$,~$\Gamma_H=0.02~\rm GeV$ and $\delta=0.007$ (see 
Eq. \ref{belpole}).
%However, the 
%aforementioned point produces 
%$\Omega h^2 \sim 0.05$ which fails to accommodate the correct 
%relic abundance data. Obtaining $\Omega h^2 \sim 0.1$ requires extreme
%fine-tuning among the parameters, especially in the singlet Higgs sector.
We also consider another simple scenario, where 
$\sigma_1$ is not the sole component of the dark matter density. For example,
if the second lightest CP-odd sneutrino eigenstates, 
$\sigma_2$ is degenerate with $\sigma_1$, then it will be present 
during the time of decoupling to contribute to the relic abundance. 
An example is depicted in column (C) of Table \ref{tabnusm} which 
possesses the desired values of the thermal relic density and $\svgg$.

%%%
\subsubsection{{\large{\it Degenerate $\phi_1$-$\sigma_1$ Dark Matter}}}
Both of these components can be present today, if their masses are 
(near) degenerate\footnote{Mass splitting $\Delta_m$ between 
$\phi_1$,$\sigma_1$ should be $\mathcal{O}(100 ~\rm keV)$, so
that the heavier state is sufficiently long-lived.}.
%Arkani Hamed on dominant left--handed state, via direct detection 
In this case, this degenerate DM annihilates through singlet-like CP
odd Higgs 
($A_s$) resonance to $\gamma\gamma$, as illustrated in Fig. \ref{fig:1}. 
Similar to the previous case, a singlet--like CP-odd Higgs, with dominant 
branching ratio to di-photon final state, helps in satisfying 
the continuum constraints; while enhancing DM annihilation
cross-section to di-photon final state in the vicinity of the narrow resonance.  
%A small width also helps in achieving the right thermal relic density.
%%%%%
\begin{figure}[h!]
\begin{center}
\includegraphics[height=2.45in]{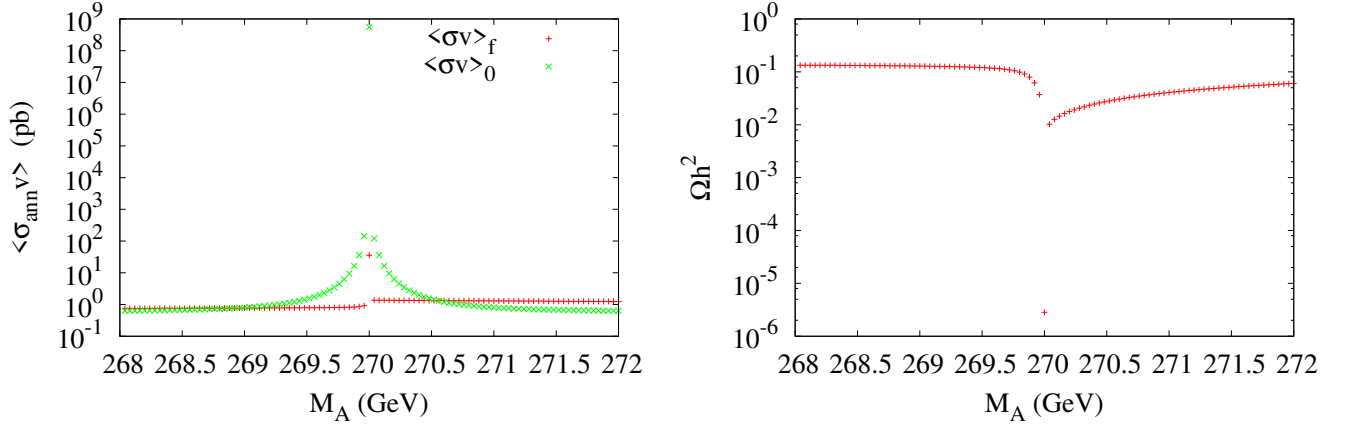}
\end{center}
\caption{Relic density of mixed $\phi_{1},~ \sigma_1$
type DM  with a mass of 135 GeV, has been
plotted against the singlet--like CP-odd Higgs pole mass. In the left panel
the thermally avaraged cross-sections at freeze-out temperature ($T_f$) and at 
late time (the speed of DM $v \sim 0.001$ or $T =T_0 \sim 10^{-4}$ GeV) have
been shown in red and green respectively. In the right panel 
the CP-odd Higgs mass is varied to demonstrate the relic abundance below 
and above the threshold (270 GeV).
}
\label{fig:odd}
\end{figure}
%%%%%%
%Relic density of mixed $\phi_{1},~ \sigma_1$
%type DM 
%around the singlet--like CP-odd Higgs pole has been plotted; 
%the CP-odd Higgs mass is varied to demonstrate the relic abundance 
%below and above the threshold (270 GeV).
To estimate $\svgg$, as mentioned, we use Eq. \ref{belpole}. 
The coupling $C$ of the singlet--like CP-odd Higgs ($A_s$) 
with $\sigma_1-\phi_1$, in the gauge eigen-basis, is given by, 
\begin{equation}
C=-2 \kappa  v_s y_r Z^A_{s3} Z^I_{12} Z^R_{12} -
\sqrt{2} Z^A_{s3} T_r Z^I_{12} Z^R_{12};
\label{Degcoupling}
\end{equation}
where, $Z^A$ denotes the mixing matrices of the CP-odd Higgs and
all other symbols are as defined before. 

As in the previous case, $\svgg$ can easily be enhanced by increasing $C$ 
(thus $y_r$ and/or $\lambda$), but that could lead to small relic abundance. 
A light higgsino--like $\chi^{\pm}$ is desired
to enhance the same. Again, we consider $m_{A_s}<2m_{\phi_1-\sigma_1}$, i.e.,
part of the parameter space where the pole of the propagator is below the threshold, to 
obtain the right thermal relic abundance. In particular, here we require 
extremely singlet like pseudo-scalar Higgs so that Br($A_s\rightarrow \gamma
\gamma$) (see Eq. \ref{belpole}) 
can be very large to account for the desired $\svgg$. Also,  
the coupling $C$ should be small so that the correct relic abundance can be obtained
through the processes mediated by the other Higgs states. This in turn prefers
small values of $y_r$ if one does not want to have unnatural cancellation
among different parameters in Eq. \ref{Degcoupling}.
However, here for $\delta >0$ (see Sec. \ref{bwr}), below the threshold, in our
region of interest, we find 
$A_s$ dominantly decays to a pair of $\nu_R$ which can substancially enhance
the decay width of $A_s$. It also affects the DM density as  
DM can also annihilate in to a pair of $\nu_R$ via $A_s$ resonance leading 
to a small relic density. To circumvent this problem, we add an additional
Majorana mass term ($m_{N}\nu_R \nu_R$) to the Lagrangian which enhance the
masses for sterile neutrinos to kinematically forbids this decay channel. Thus
the pole can appear very close to the threshold and $\svgg$ can be 
extremely large in this region (as the branching fraction $A_s\rightarrow \gamma
\gamma$ can be as large as $\sim$70\%).

To demonstrate our results, we present in the left panel of Fig. \ref{fig:odd},
the thermally averaged cross-section, when the pole falls above the threshold
(270 GeV), is a little larger than the same when the pole falls below the
threshold. Also, as the right panel of Fig. \ref{fig:odd} shows,
in the former case, freeze-out happens a little later. Consequently, 
the thermal relic abundance decreases in the former case. Note that, due to the
much smaller width of $A_s$ compared to that of $H_2$ in the previous case,
the effect of the resonance on the relic density is quite small.  
%Annihilation channels mediated by the light Higgs contributes significantly 
% to the relic density too. 
%For $\delta >0$ (see Sec. \ref{bwr}), below the threshold, $A_s$ dominantly 
%decays to a pair of $\nu_R$. 
%The total width, in this region, therefore, can be
%reduced by reducing $y_r$. Note that, this, in turn, reduces the effective 
%Majorana mass ($y_r v_s$) for $\nu_R$, while also affecting $\svgg$ and 
%the relic density. 
%Adding Majorana mass term for $\nu_R$, however, can 
%kinematically forbid this decay channel. Consequently, the same forbids 
%DM annihilation to a pair of $\nu_R$ via $A_s$ resonance, thus 
%increasing the relic density. 
 
% Thus the pole can appear very close to the threshold and $\svgg$ can be 
%extremely large in this region, as the branching fraction $A_s\rightarrow \gamma
%\gamma$ can be as large as $\sim$70\%. 
%{( good to specify $\gamma$ and $\delta$)}

A benchmark point in column (A) of Table \ref{tabnusm} has been presented
to summarize the results. The relic density can be obtained mainly via the
annihilation channels mediated via the off-shell CP-even Higgs bosons. In the
calculation of $\svgg$, we have $\rm Br(A_s \rightarrow \gamma \gamma)
\simeq 5 \times 10^{-4}$,~$\Gamma_{A_s}=0.003\rm ~GeV$ and $\delta=0.002$.
All phenomenological constraints, from both direct and indirect detection 
data, can easily be satisfied for the benchmark point (A) shown
in Table \ref{tabnusm}, thanks to the singlet nature of $A_s$. 

%----------------------------------------------------------------------------

%-----------------------------------------------------------------------------
\subsubsection{\large{\it Constraints from direct detection}}
An interesting issue for the models, which give the desired $\svgg$, 
is to address constraints coming from the spin-independent direct detection, 
particularly in the light of XENON-100 and LUX data. 
In the NMSSM, the parameter 
space for a 130-135 GeV neutralino DM, achieving the desired $\svgg$, is 
constrained by the present bound $\sigma(p)_{SI} \lsim 1.2 \times 10^{-8}$~pb
and $\sigma_{SI} \lsim 1.5 \times 10^{-9}$~pb for $M_{DM} \sim 130-135$~GeV~from
XENON-100 \cite{Aprile:2012nq} and LUX \cite{LUX13} respectively. An important
issue concerns the quark coefficient in the nucleon which may lead to large
theoretical uncertainty in the calculation. In this work, we always
keep the default values that are used in \texttt{micrOMEGAs-3}. 

Being a real scalar, $\phi_1$ ($\sigma_1$) interacts with nucleons via 
Higgs exchange processes. As the Yukawa coupling is very small 
($\sim 10^{-7}$), the coupling $\phi_1/\sigma_1-H-\phi_1/\sigma_1$ is
principally determined by $T_r S\tilde{\nu}^c_R\tilde{\nu}^c_R$ and 
$\kappa v_s S^{*}\tilde{\nu}^c_R\tilde{\nu}^c_R$, ~$\lambda y_r v_{u,d}
H_{d,u}^{*} \tilde{\nu}^c_R\tilde{\nu}^c_R$ terms, arising from the
soft-breaking sector and the F-term contributions respectively. 
Singlet--like Higgs does not couple to the quarks or gluons at tree--level, 
while the coupling of the doublet--like Higgs with $\tilde{\nu}^c_R
\tilde{\nu}^c_R$ is quite small. Also, since both $\phi_1$ and $\sigma_1$ 
are dominantly right-chiral, $Z$ boson exchange does not lead to large
$\sigma(p)_{SI}$ even when these are degenerate. The only term, 
which gives rise to dominant contribution to $\sigma(p)_{SI}$ originates 
from the F-term contribution $(\lambda \dfrac{y_r}{2} \tilde{\nu}^c_R
\tilde{\nu}^c_R H_u.H_d+\text{h.c})$. While both $\lambda$ and $y_r$ appears in
$\svgg$,
note that it is possible to achieve large $\svgg$ with small $y_r$ by 
increasing the soft-term $T_r$ appropriately. 
In summary, the scenarios proposed 
here are not significantly constrained by the XENON-100, these are 
moderately constrained by the LUX data. 

%Comment on singlet components in \sigma_1/\phi_1, the mixing is 
%constrained by relic abundance and direct+indeirect searches; 
%Comment on the singlet nature of the intermediate particle, how much  
%can it be relaxed in the present scenario. 
%Comment on how close can be MA/MH and (2*m, m= mass of the DM). How to obtain
%correct relic and correct photon signal, when the difference $4m^2 \delta$ is 
%varied.
%Comment on the corrections on the neutrino mass from the mass-splitting
%of heavier CP-even and CP-odd (~400 GeV) left-handed neutrinos.
%g_\mu -2 
%tri-lepton sector--> bound on chargino mass  

\section{Summary and Conclusions}
\label{conclusion}
We have explored the possibility that the annihilation 
of 130-135 GeV right-chiral sneutrino DM into two photons can produce 
the observed line-like feature in the Fermi-LAT data. 
In this context, we examine the candidature of right-sneutrino dark matter -- a scenario
that can have somewhat unusual phenomenological implications. It is, however, seen that
the augmentation of the MSSM just with right-chiral neutrino superfields is inadequate.
The difficulty arises from severe
constraints on various annihilation channels of the dark matter, most notably 
into $Z H$ and $b \bar{b}$, derived from the continuum flux of photons.
However, in the extension of the next-to-minimal model (NMSSM),
annihilating right-chiral sneutrino DM, can produce the observed line feature. 
Due to the extra singlet field present in this model, a singlet--like CP-odd or CP-even 
Higgs boson resonance produces adequate annihilation cross-section to fit the observation. 
We find that in case of a CP-odd Higgs resonance, one needs the lightest CP-even and the 
lightest CP-odd right-chiral sneutrino states to be (almost) degenerate. In  the latter case 
however, this is not a requirement. We present a few benchmark points to substantiate our 
claims and highlight the spectrum which is consistent with the data.  While our benchmark 
points also satisfy the present direct detection bounds, improved bounds in near
future may be able to explore the viability of our scenario. In addition, we show that 
when the pole in the resonance is a little below twice the mass of the DM, the
thermal production of right-chiral sneutrino dark matter can be sufficient to 
also account for the DM abundance, as required by the CMBR data, especially
in the case of degeneracy in the sneutrino sector. 
 
\section{Acknowledgement}
AC and DD would like to thank Florian Staub for useful discussions about 
the code \texttt{SARAH}. 
The work of AC,  BM and SKR was partially supported by funding available 
from the Department of Atomic Energy, Government of India, for the Regional
Centre for Accelerator-based Particle Physics (RECAPP), Harish-Chandra Research
Institute (HRI). DD acknowledges support received from the DFG, project no.\
PO-1337/3-1 at the Universit{\"a}t W{\"u}rzburg. 
DD thanks RECAPP, Allahabad, for hospitality during the initial part of the 
proejct.

% add generation index i in Yukawas and soft terms
%*********************************************************************

\end{document}